\documentclass[12pt]{article}

\usepackage{arxiv}

\usepackage[utf8]{inputenc} 
\usepackage[T1]{fontenc}    
\usepackage{hyperref}       
\usepackage{url}            
\usepackage{booktabs}       
\usepackage{amsfonts}       
\usepackage{nicefrac}       
\usepackage{microtype}      
\usepackage{lipsum}
\usepackage{comment}
\usepackage{subcaption}
\usepackage{graphicx}
\usepackage{soul}           
\usepackage{amsmath}
\usepackage{bm}
\usepackage[sorting=none,giveninits,style=numeric-comp,doi=false,isbn=false,url=false,date=year,eprint=false,maxnames=6]{biblatex}
\usepackage{chngcntr}
\usepackage[font={small}]{caption}

\DeclareMathOperator*{\argmax}{argmax} 

\title{Data-Driven Collective Variables\\for Enhanced Sampling}

\author{
  Luigi Bonati\\
  Department of Physics, ETH Zurich, 8092 Zurich, Switzerland\\
  and Institute of Computational Sciences, Universit{\`a} della Svizzera italiana (USI),\\via Buffi 13, 6900 Lugano, Switzerland \\
   \And
  Valerio Rizzi\\
  Department of Chemistry and Applied Biosciences, ETH Zurich, 8092 Zurich, Switzerland\\
  and Institute of Computational Sciences, USI, via Buffi 13, 6900 Lugano, Switzerland \\
   \AND
  Michele Parrinello\\
  Department of Chemistry and Applied Biosciences, ETH Zurich, 8092 Zurich, Switzerland,\\
  Institute of Computational Sciences, USI,via Buffi 13, 6900 Lugano, Switzerland, \\
  and Italian Institute of Technology, Via Morego 30, 16163 Genova, Italy
} 

\begin{document}
\maketitle

\begin{abstract}
Designing an appropriate set of collective variables is crucial to the success of several enhanced sampling methods. Here we focus on how to obtain such variables from information limited to the metastable states. We characterize these states by a large set of descriptors and employ neural networks to compress this information in a lower-dimensional space, using Fisher's linear discriminant as an objective function to maximize the discriminative power of the network.
We test this method on alanine dipeptide, using the non-linearly separable dataset composed by atomic distances. We then study an intermolecular aldol reaction characterized by a concerted mechanism. 
The resulting variables are able to promote sampling by drawing non-linear paths in the physical space connecting the fluctuations between metastable basins.
Lastly, we interpret the behavior of the neural network by studying its relation to the physical variables. Through the identification of its most relevant features, we are able to gain chemical insight into the process.
\begin{figure}[h!]
  \centering
  \includegraphics[width=0.65 \columnwidth]{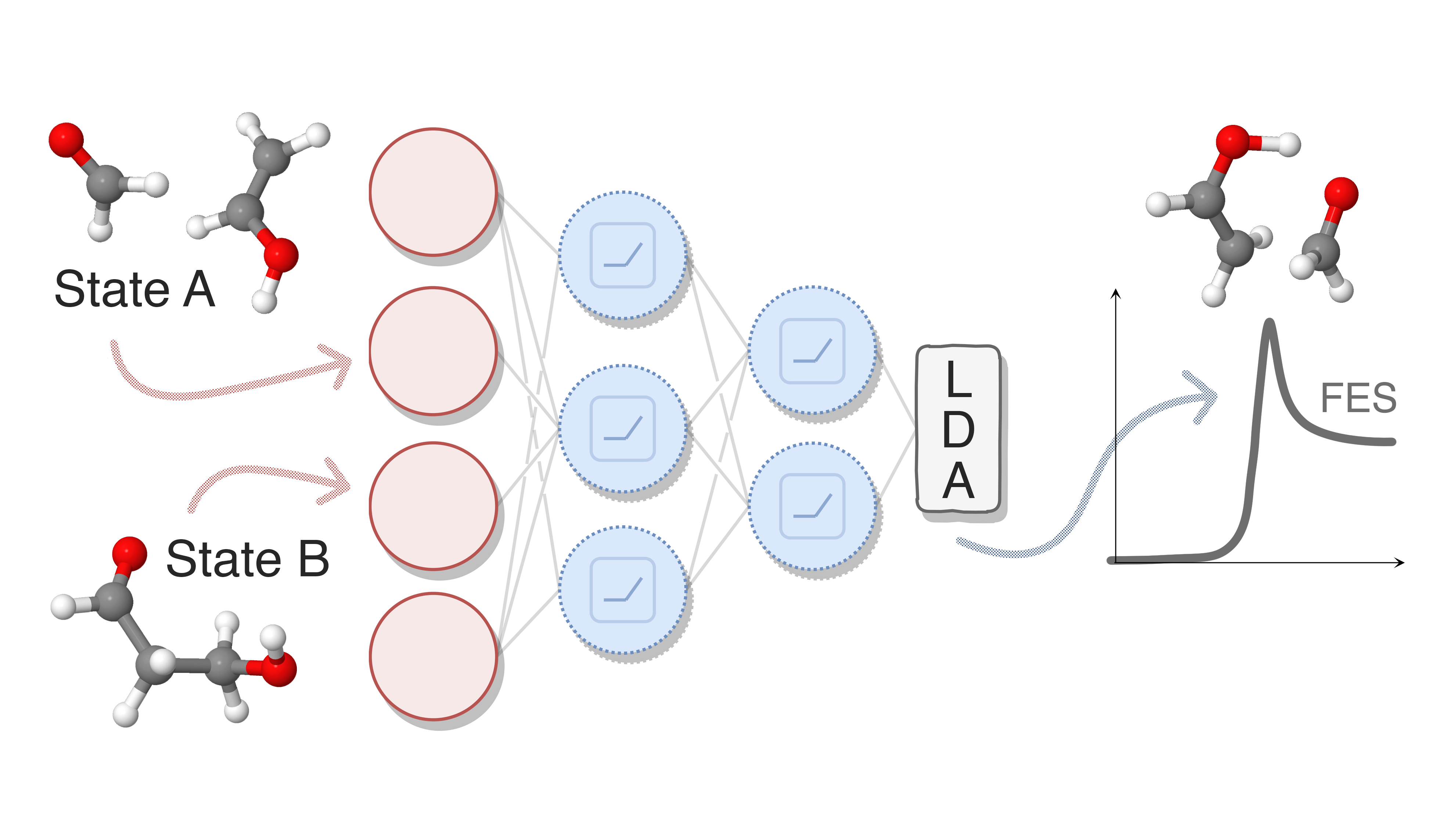}
\end{figure}
\end{abstract}

\clearpage

In recent decades, enhanced sampling methods have become one of the main tools of computational science. They have extended the scope of atomistic simulations, allowing for long time scale phenomena to be simulated. Starting from the pioneering work of Torrie and Valleau \cite{Torrie1977}, a large class of such methods relies on the introduction of collective variables (CVs) defined as functions of the atomic coordinates. Once these variables are identified, a bias potential that is a function of the selected CVs is added to the interaction potential to accelerate sampling \cite{Valsson2016}. 
The CVs are chosen so as to describe the hard-to-sample modes of the system, but this is non-trivial, especially when the system under investigation is complex and its transition pathways are unknown.
Not surprisingly, much effort has been devoted to the identification and improvement of useful CVs. Very recently, machine learning methods of varying complexity have been also used to this effect \cite{Ma2005,Ribeiro2018,Chen2018,Schoberl2019a,Rogal2019,Ravindra2019,Chen2019,Wehmeyer2018b,Hernandez2018}. Building on our previous studies, we introduce a new method that uses deep neural networks (NNs) to perform a nonlinear featurization of a large set of input descriptors in order to build effective CVs. However, before describing our approach we recall some recent findings that give a clue to what we plan to do.

The free energy landscape of physical systems can be described as being made up of islands of metastability in a sea of improbable configurations. The metastable states are connected by narrow passageways that allow rare but crucial transitions from one state to another to take place. Very recently our group and others \cite{Mendels2018,Sultan2018} have developed a family of efficient CVs based only on the fluctuations in the metastable basins. This defies the conventional wisdom that for a variable to be effective, it must contain explicit information about the whole reaction path or at least the transition state \cite{Ma2005,Peters2017}. 

As an example of this approach, we briefly recall how the method called harmonic linear discriminant analysis (HLDA) \cite{Mendels2018} works. One first identifies a small set of descriptors capable of discriminating between the states. The expectation value of the descriptors and their covariance matrices are computed from short unbiased runs in the two basins. With the knowledge of only these quantities, and by using a variant of the classification method that goes under the name of linear discriminant analysis (LDA)\cite{welling2005}, one obtains CVs that are linear combinations of the input descriptors.

This simple approach has proven to be highly effective in a variety of cases, such as chemical reactions \cite{Piccini2018,Rizzi2019}, crystallization \cite{Zhang2019}, ligand unbinding \cite{Capelli2019} and small peptides folding \cite{Mendels2018a}, but it has limitations. At first, the states must be linearly separated in the descriptors space, as the HLDA CV is built upon their linear combination. 
Thus, HLDA crucially relies on the identification of a small set of uncorrelated descriptors. While this is a far less demanding task than finding CVs, it requires knowledge of the system and physical intuition. This intuition can be of great help in leading to an accelerated sampling, but, sometimes it might reflect more our prejudice than the actual system behavior, possibly preventing the exploration of some of the relevant transition pathways. With the use of an appropriately designed neural network, we want to lift these limitations.

To achieve this result, we employ a NN to perform a non-linear transformation of the inputs, optimized by Fisher's linear discriminant as objective function.
The idea of improving linear techniques by combining them with NNs has been applied to LDA for classification purposes \cite{Dorfer2016} and to other problems in different contexts\cite{Andrew2013,Mardt2018,Chen2019,Hernandez2018}. Here we want to apply it to the design of CVs.

\textbf{Method} 
LDA was first introduced as a classification method. In this context, one searches for the linear combination of the input features that best separates the data in given classes. Let us consider a set of $N$ data points $\mathbf{x}_1,...,\mathbf{x}_N$ (observations) of dimension $d$ (local descriptors) belonging to $C$ classes (metastable states). The covariance matrix over all these samples can be decomposed in two terms, the so-called within class $\mathbf{S}_w$ and between class $\mathbf{S}_b$ scatter matrices. The former takes into account the fluctuations inside the basins, and corresponds to the average of the class covariances $\mathbf{S}_i$:
\begin{equation}
    \label{eq:sw}
    \mathbf{S}_w = \frac{1}{C} \sum_i ^C \mathbf{S}_i.
\end{equation}
On the other hand, $\mathbf{S}_b$ is defined as the covariance of the class means $\bm{\mu}_i$, and thus measures the fluctuations between classes:
\begin{equation}
    \mathbf{S}_b = \frac{1}{C} \sum_i ^C (\bm{\mu}_i -\bm{\mu})(\bm{\mu}_i -\bm{\mu})^T
\end{equation}
where $\bm{\mu}$ is the average of the $C$ class means.
Fisher's criterion seeks a linear projection $\mathbf{W}$ into a $C-1$ dimensional space, such that the samples show a high variance between classes and a low variance within. This is achieved by maximizing Fisher's ratio
\begin{equation}
    \argmax_\mathbf{W} \frac{\mathbf{W} \mathbf{S}_b \mathbf{W}^T}{\mathbf{W} \mathbf{S}_w \mathbf{W}^T},
    \label{eq:fisher}
\end{equation}
that measures the degree of separation between the classes.

In order to find the combination $\mathbf{W}$ which maximizes eq.~\ref{eq:fisher}, one has to solve the generalized eigenvalue problem:
\begin{equation}
    \mathbf{S}_b \mathbf{w}_i = v_i \mathbf{S}_w \mathbf{w}_i \hspace{5em}   \forall\ i=1,\dots,d
    \label{eq:eig-problem}
\end{equation}
where the eigenvectors $\mathbf{w}_i$ form the projection matrix $\mathbf{W}$ and the eigenvalues $v_i$ quantify the separation in the corresponding directions. We then obtain the compressed representations $s_i$ with $i=1,...,C-1$ by projecting the data points $\mathbf{x}$ along the corresponding eigenvectors as: $s_i = \mathbf{w}_i^T \mathbf{x}$.
\begin{figure}[t!]
  \centering
  \includegraphics[width=.9\columnwidth]{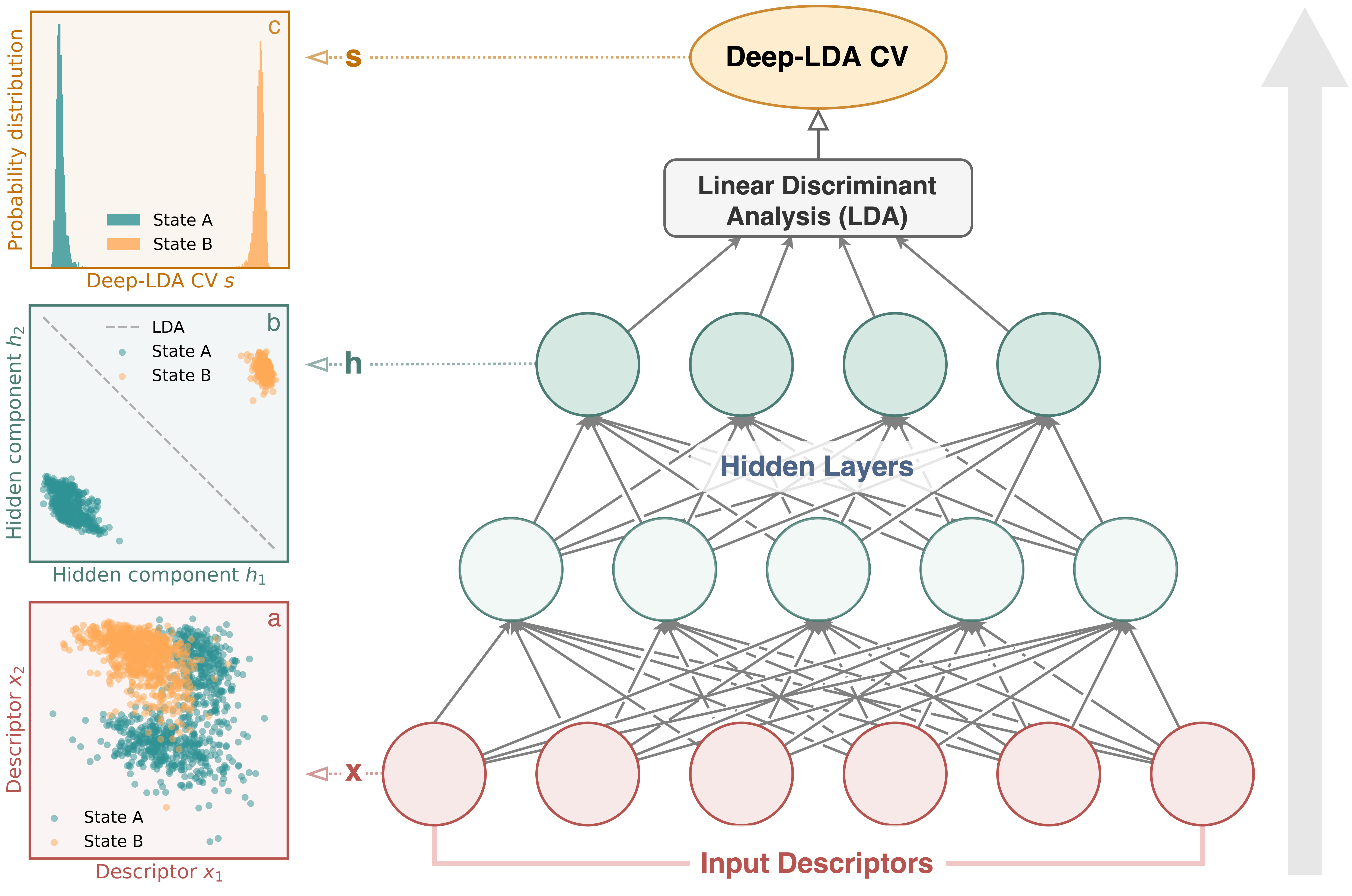}
  \caption{Scheme of the construction of the Deep-LDA CV. A set of physical descriptors are used as inputs of a feed-forward neural network. A non-linear transformation is made by the NN via the composition of several hidden layers. In the the last layer, a Linear Discriminant Analysis is performed, the direction of maximal separation between classes is determined and the CV is obtained. To better illustrate the workings of Deep-LDA, we report three panels on the left hand side of the figure. In (a), we show the distribution of a pair of typical input descriptors and note how they do not distinguish state A from state B. In (b), we plot the distribution of a pair of variables from the topmost hidden layer, along with the LDA boundary projected in this space. At this stage, after the non-linear transformation, state A and B are linearly separable. In (c), we report the probability distribution of the Deep-LDA CV for the two states. }
  \label{fig:method}
\end{figure}
By choosing $s_i$ as a set of CVs, LDA has been used in enhanced sampling applications. In previous work, a harmonic average of the covariance matrices was performed in eq.~\ref{eq:sw}, hence the name HLDA \cite{Mendels2018}. However, in this paper we use the standard LDA version.

We now illustrate how LDA can be combined with a neural network (see fig.~\ref{fig:method}). We feed a number of descriptors to the NN that reduces the dimensionality of the data through a succession of continuous non-linear transformations represented by $f_\theta(\mathbf{x})$ with parameters $\theta$. 
We then perform LDA on the topmost hidden space representation $\mathbf{h}=f_\theta(\mathbf{x})$, and use the eigenvalues of eq.~\ref{eq:eig-problem} to optimize the NN.
From now on, we restrict the discussion to the typical case of two-classes, where the loss function reads:
\begin{equation}
    \mathcal{L}=-v_1
    \label{eq:loss}
\end{equation}
and $v_1$ is the only non-zero eigenvalue of eq.~\ref{eq:eig-problem}.
By maximizing the LDA eigenvalue, one increases the discriminative power of the NN and learns at the same time linearly separable latent features (see fig.~\ref{fig:method}b). The resulting Deep-LDA CV, reported in fig.~\ref{fig:method}c, is obtained by projecting the output of the NN $\mathbf{h}$ along the LDA eigenvector $\mathbf{w}_1$ as follows: $s = \mathbf{w}_1^T \mathbf{h}$. Generalizing the loss function to multi-state problems is simple and can be done either by maximizing the smallest of the $C-1$ eigenvalues or their sum \cite{Dorfer2016}.

To make the optimization more stable we regularize the within scatter matrix by adding a multiple of the identity matrix
\begin{equation}
    \mathbf{S'}_w=\mathbf{S}_w+\lambda\mathbf{I},
    \label{eq:lambda}
\end{equation}
as done in ref.~\cite{Dorfer2016}.
Furthermore, we transform the generalized eigenvalue problem into a standard one by performing a Cholesky decomposition of $\mathbf{S'}_w = \mathbf{L}\ \mathbf{L}^T$ \cite{welling2005}. By simple manipulation, we rewrite eq.~\ref{eq:eig-problem} as:
\begin{equation}
    \widetilde{\mathbf{S}}\ \widetilde{\mathbf{w}}_i = v_i\  \widetilde{\mathbf{w}}_i \hspace{4em}   i=1,2
\end{equation}
where $\widetilde{\mathbf{S}}=\mathbf{L}^{-1} \mathbf{S}_b (\mathbf{L}^T)^{-1}$ and $\widetilde{\mathbf{w}}_i=\mathbf{L}^T \mathbf{w}_i$. In this way, we have a symmetric eigenvalue problem which can be solved using the Pytorch library \cite{AdamPaszke;SamGross;etal2017}. This allows training Deep-LDA networks with backpropagation and mini-batch gradient descent.

We note that, in the context of classification, eq.~\ref{eq:loss} is used as loss function and training is stopped when the configurations are correctly labeled \cite{Dorfer2016}. However, during the optimization process, the separation between the projected classes of eq.~\ref{eq:fisher} tends to increase without bounds either by enlarging the distance between classes or by collapsing them into delta-like distributions. Since we intend to use the compressed representation in enhanced sampling methods, neither very large distances between states nor too narrow basins are suitable for our purpose.
We regulate the width of the projected classes with $\lambda$ (eq. \ref{eq:lambda}), which provides a lower bound of the within class covariances, and we control the distance by modifying the loss function as follows: 
\begin{equation}
    \mathcal{L} = -v_1 - \frac{\alpha}{1+\left( \overline{s^2}  -1 \right)^2}
    \label{eq:lorentzian}
\end{equation}
where $\alpha$ represents the magnitude of the regularization (see Supporting Information SI-1) and $\overline{s^2} = \frac{1}{n_b}\sum_{j=1} ^{n_b} s^2_j$ is the average of the squared CV over a batch of $n_b$ configurations.
The second term in eq.~\ref{eq:lorentzian} is maximum for $\overline{s^2}=1$, thus it acts as an effective constraint on the distance between the basins. More details about the loss function and the optimization procedure are reported in the Supporting Information (SI).

After the training reaches convergence, we use the Deep-LDA CV in combination with enhanced sampling methods. In principle, this could be done with any method, such as Metadynamics \cite{Laio2002} or Variational Enhanced Sampling \cite{Valsson2014,Bonati17641}. In this work, we choose to employ a recently developed evolution of Metadynamics, called on-the-fly probability enhanced sampling (OPES) \cite{Invernizzi2019}, which builds the bias from an on-the-fly estimation of the probability distribution.

In OPES, the probability distribution at iteration $n$ is given by 
\begin{equation}
    P_n(\mathbf{s})=\frac{\sum_k ^n w_k G(\mathbf{s},\mathbf{s}_k)}{ \sum_k ^n w_k },
\end{equation}
where $G(\mathbf{s},\mathbf{s}_k)$ is a multivariate Gaussian and the weights are computed from the previously deposited bias potential $w_k=e^{\beta V_{k-1}(\mathbf{s}_k)}$, with $\beta$ the inverse temperature.
In turn, the bias potential is defined as:
\begin{equation}
    V_n(\mathbf{s}) = \left(1-\frac{1}{\gamma}\right)\frac{1}{\beta} \log \left( \frac{P_n(\mathbf{s})}{Z_n}+\epsilon \right) %
    \label{eq:bias}
\end{equation}
where $Z_n$ is a normalization factor and the bias factor $\gamma$ is a parameter, that, as in Well-Tempered Metadynamics \cite{Barducci2008}, determines the broadening of the biased distribution.
Finally, $\epsilon$ is a regularization term, that sets a limit to the maximum value of the bias, thus limiting the exploration of higher free energy regions.

Once the iterative process converged, the free energy surface (FES) can be computed as: 
\begin{equation}
    F_n(\mathbf{s})=-\frac{1}{\beta}\log P_n (\mathbf{s})
    \label{eq:fes}
\end{equation}
and simple umbrella-sampling reweighting gives access to all the static properties of interest. We refer the reader to ref.~\cite{Invernizzi2019} for further details.

The combination of Deep-LDA and OPES presents a few noteworthy features. Enhancing the dynamics along the Deep-LDA CV typically spreads the action of the bias over a fairly large number of degrees of freedom. This helps greatly in promoting transitions but also increases the risk of exploring unwanted regions of the phase space. Therefore, the capability of OPES to restrict the exploration by choosing appropriately the $\epsilon$ parameter in eq.~\ref{eq:bias} helps in focusing the sampling. Other benefits of OPES include its robustness, due to a small number of free parameters and a convergence faster than Metadynamics.

Before proceeding to the applications, we summarize the method for clarity:
\begin{enumerate}
    \item Run short unbiased MD runs in the metastable states and compute the descriptors
    \item Construct a CV by training a NN with LDA as the objective function
    \item Use the Deep-LDA CV to enhance the sampling and obtain the FES
\end{enumerate}

\clearpage
\textbf{Alanine dipeptide}
The first example that we choose is alanine dipeptide, a small molecule often used to benchmark sampling methods. Alanine has two metastable states that are well described using the pair of Ramachandran angles $\phi$ and $\psi$, which forms a nearly ideal set of CVs. 

However, since here we wish to show the strength of the method, we deliberately put ourselves in a situation more complex than necessary and choose a general set of descriptors. 
This is more similar to the situation one encounters in practice, where optimal CVs are difficult to find.
The key here is Deep-LDA's ability to handle a large number of descriptors. Unfortunately, we cannot directly use the atomic coordinates, since the resulting CV would not be rotationally and translationally invariant.
A natural choice is to use coordinate combinations that are invariant under such symmetries, for instance distances, angles and dihedrals.
To make things even more challenging, we choose only distance-based descriptors. In such a base the states cannot be linearly separated.
Instead Deep-LDA, which is intrinsically non-linear, overcomes this difficulty and leads to satisfactory results.

We run short unbiased trajectories in the two basins, using as descriptors the 45 distances between heavy atoms, and with the data thus accumulated we train a Deep-LDA CV (see SI-3). We developed an interface to load the Pytorch model in the open-source plug-in PLUMED2 \cite{Tribello2014} that allows using the Deep-LDA CV with enhanced sampling methods. The code and the input files needed to reproduce the simulations are openly available in the PLUMED-NEST repository with plumID:20.004.

\begin{figure}[b!]
  \centering
  \includegraphics[width=0.5\columnwidth]{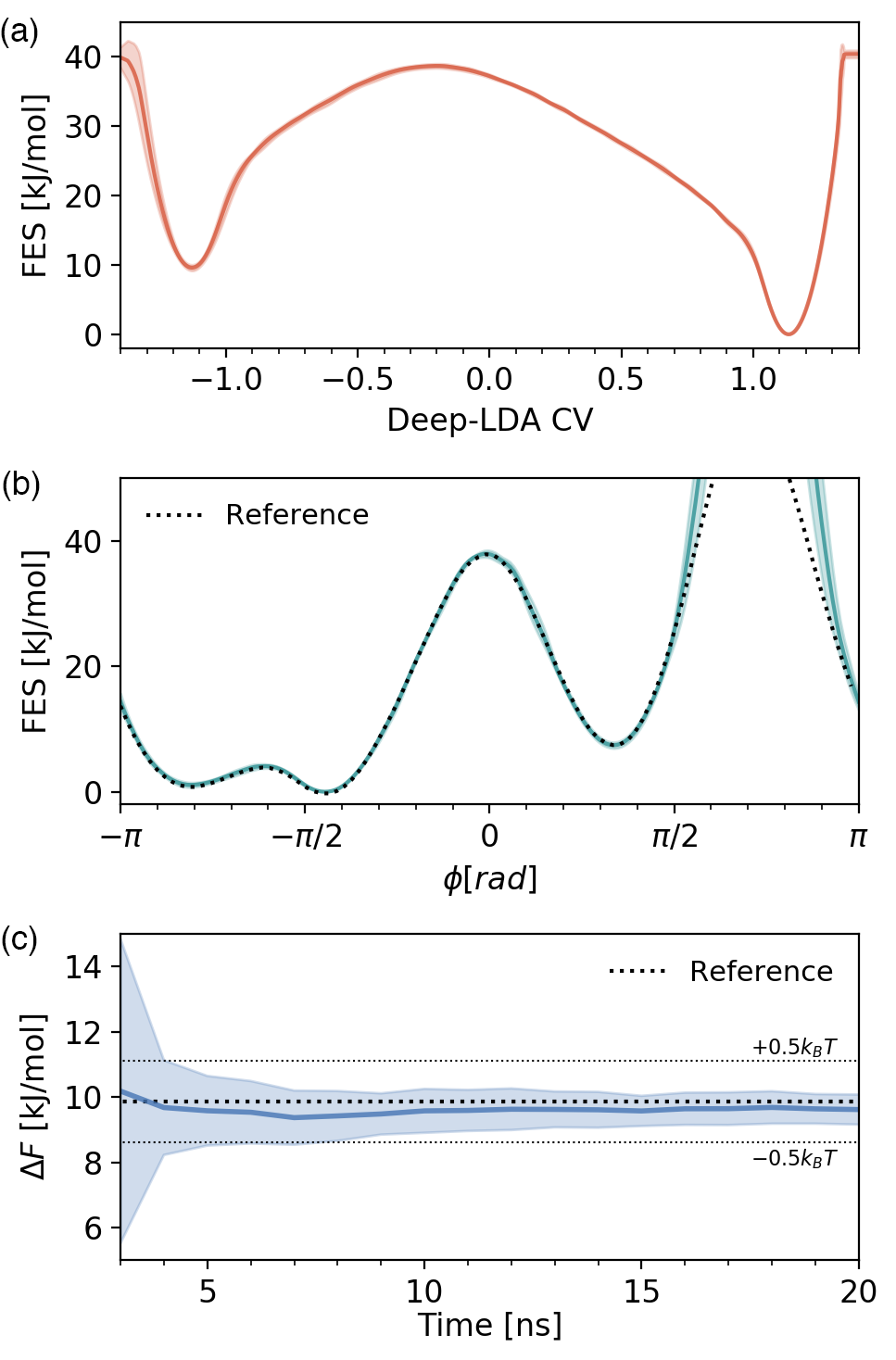}
  \caption{We show here alanine dipeptide convergence tests in which we run 10 independent simulations with OPES and the Deep-LDA CV, from which we extract the mean (solid line) and the standard deviation (shadow region).
  In (a) we report the FES along the Deep-LDA CV, calculated from eq.~\ref{eq:fes}. In (b), we show the FES along the $\phi$ angle, computed with a reweighting technique (see eq.~SI-3.1). The dotted line represents the reference obtained when biasing directly $\phi$ and $\psi$. In (c), there is the free energy difference between the two metastable states as a function of time (see eq.~SI-3.2). 
  }
  \label{fig:ala2-convergence}
\end{figure}

Enhancing the fluctuations of the Deep-LDA CV leads to a highly diffusive behavior comparable to the simulations where the pair of Ramachandran angles is biased (see fig.~SI-3.1). In fig.~\ref{fig:ala2-convergence}a we report the FES along the Deep-LDA CV. Furthermore, we compare the projection of the FES along $\phi$ with more standard calculations that use $\phi$ and $\psi$ as CVs (see fig.~\ref{fig:ala2-convergence}b).
The full landscape in terms of $\phi$ and $\psi$ is reported in fig.~SI-3.3.
We then monitor the free energy difference between the two basins as a function of time (fig.~\ref{fig:ala2-convergence}c) and observe a rapid convergence to the reference value.
These results show that Deep-LDA is able to reproduce the two dihedrals which represent the slowest degrees of freedom.

\begin{figure}[h!]
  \centering
  \includegraphics[width=0.6\columnwidth]{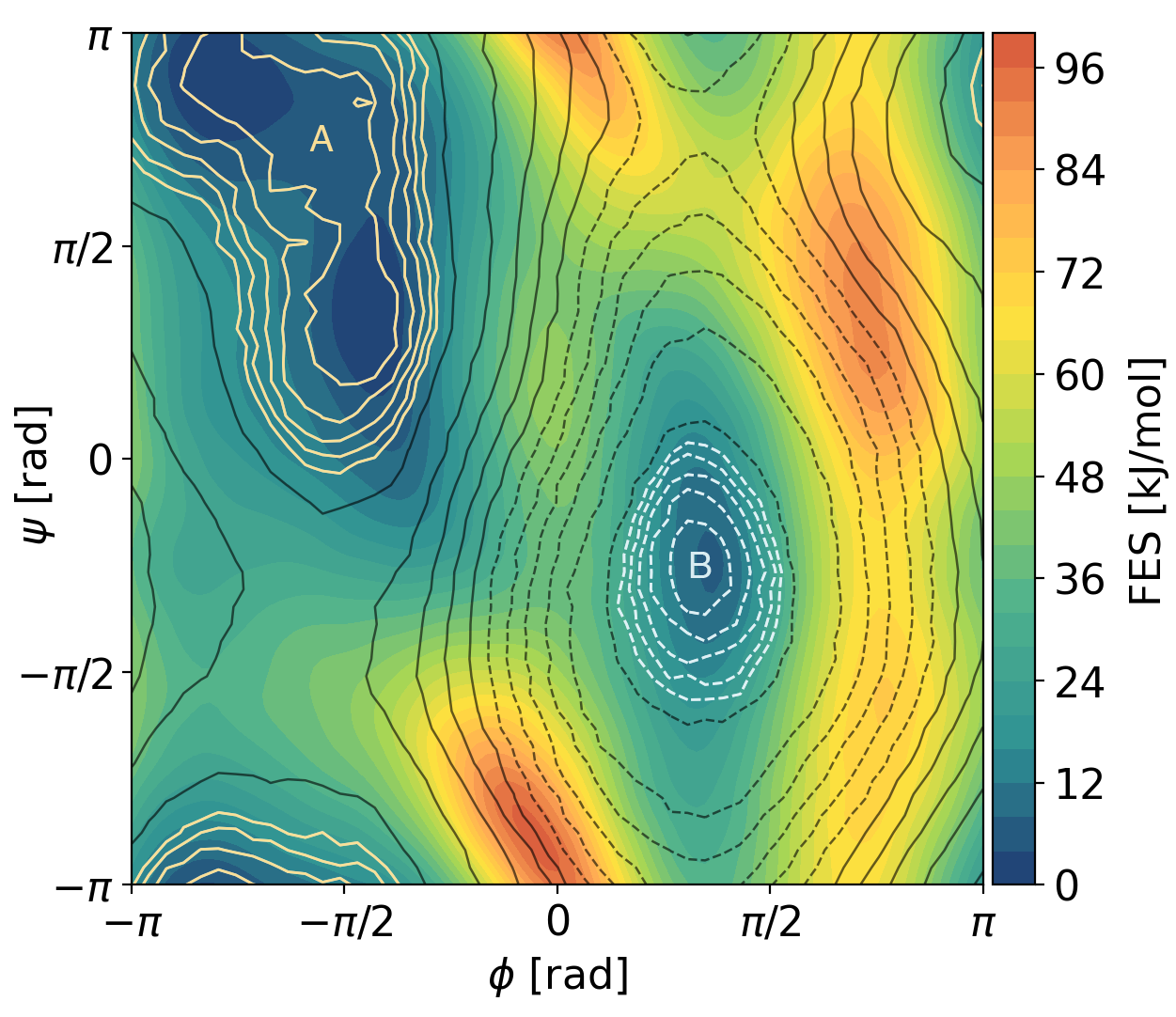}
  \caption{We show the isolines of the conditional probability of the Deep-LDA CV with respect to the Ramanchadran angles, on top of a reference free energy surface.
  The highlighted lines correspond to the regions of the two metastable states used for training, and they are spaced by 0.02. The black lines are extrapolated by the network and have a much larger spacing of 0.2. The isolines are solid or dashed depending on whether they are above or below zero.}
  \label{fig:ala2-fes}
\end{figure}

To illustrate this point further, we present in fig.~\ref{fig:ala2-fes} the isolines of the Deep-LDA CV on top of the FES, both projected onto $\phi$ and $\psi$ (see also fig. SI-3.2).
As there is no direct correspondence between the dihedral angles and the CV $s$, the isolines have been computed from the conditional probability $p(s\,|\,\phi,\psi)$ in the statistical ensemble generated by an OPES calculation with a uniform target distribution in the $(\phi,\psi)$ space.
Note that the training is performed using data from the two basins alone and the isolines corresponding to the intermediate regions are extrapolated by the network.
When the Deep-LDA fluctuations are enhanced by OPES, the system is driven along the direction perpendicular to the isolines of the CV. It is remarkable how well this direction is correlated to the low free energy path that connects the two basins. 

Similar results were obtained with different NN architectures and regularization parameters, as well as by using Well-Tempered Metadynamics (see fig.~SI-3.5), demonstrating the robustness of the method on the choice of parameters and the enhanced sampling algorithm used.

\textbf{Aldol reaction}
The second example is the aldol reaction between vinyl alcohol and formaldehyde. This reaction presents several pathways and products \cite{Yang2017}. Here, for simplicity, we focus on the most probable one, which has a  barrier of $\sim 150$ kJ/mol, as estimated in static calculations. This represents a challenging test since it is not obvious whether the reaction's concerted mechanism can be captured with information coming exclusively from unbiased simulations of reactants and products. 
Once again, in the spirit of performing the computation in a blind manner, we build the descriptor set out of all the interatomic distances. We rely on the ability of Deep-LDA to combine them in a meaningful and optimal way and to point out the most relevant ones.
Previous experience \cite{Rizzi2019} has shown that a good input set for chemical reactions can be obtained from the interatomic distances $r$, using the following contact function:
\begin{equation}
    c_{ij}(r)=\frac{ 1 - \left(\frac{ r }{ \sigma_{ij} }\right)^{n} }{ 1 - \left(\frac{ r }{ \sigma_{ij} }\right)^{m} }
    \label{eq:contacts}
\end{equation}
where $\sigma_{ij}$ are typical bonding lengths between atoms of species $i$ and $j$. 

This non-linear pre-processing of the distances is crucial in the case of linear methods. In the present context, this step is not strictly necessary, thanks to the non-linearity provided by the NN. Nevertheless, it can be of help in focusing on the relevant degrees of freedom. As in the case of NN based interatomic potentials \cite{Carleo2019}, the use of chemically informed descriptors facilitates learning from the data. In fig.~SI-4.4 we show a simulation employing the distances as inputs of the Deep-LDA CV.

\begin{figure}[b!]
  \centering
  \includegraphics[width=1\columnwidth]{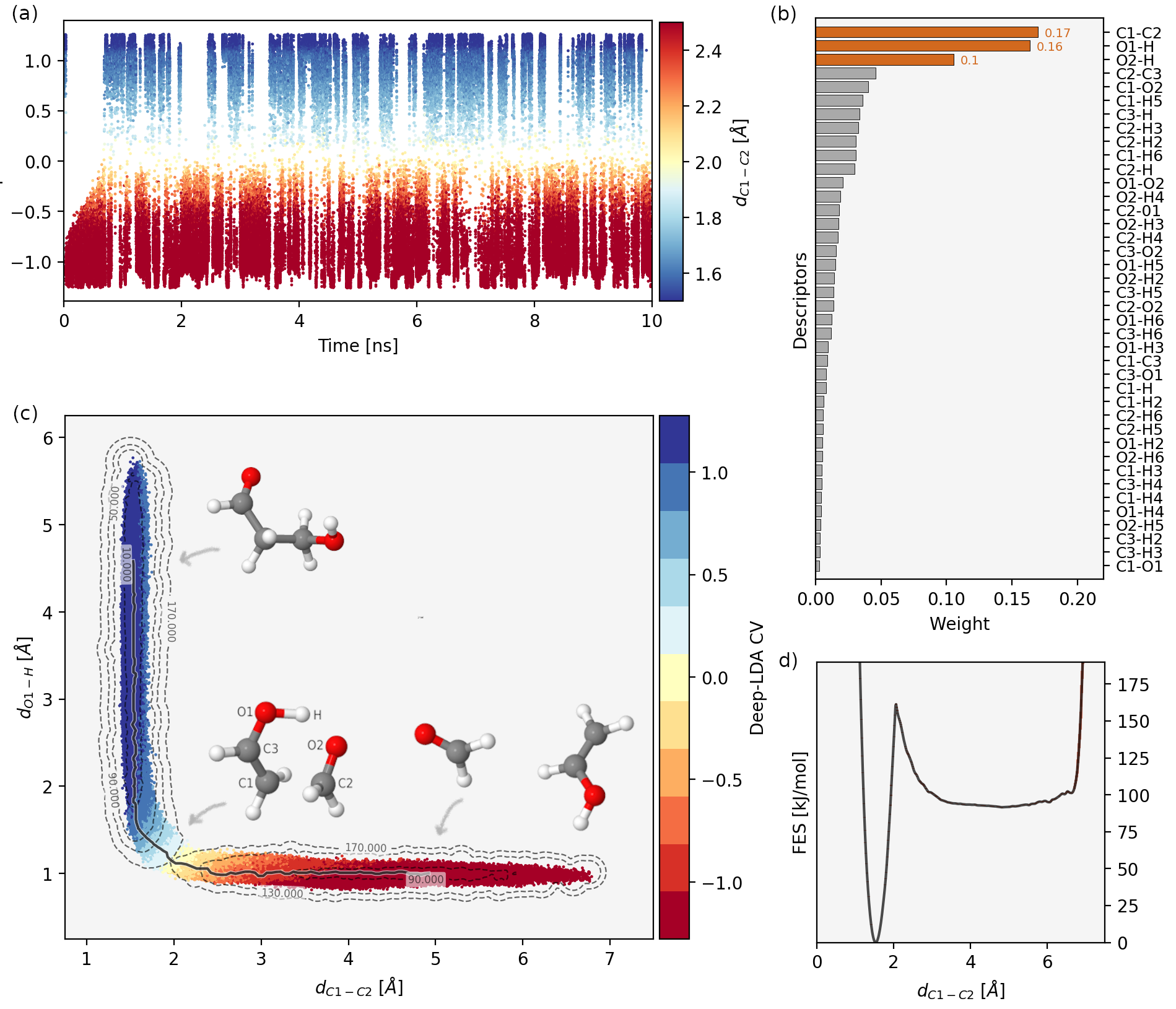}
  \caption{Results of the OPES simulation when enhancing the Deep-LDA CV. (a) Time evolution of the Deep-LDA CV. The points are colored according to the C1-C2 distance. (b) Features importance analysis, according to the magnitude of the weights of the first layer. The rankings are normalized so that their sum is equal to one. The first three inputs, separated by a gap from the following ones, are colored in orange. (c) Distribution of the visited configurations in the plane of the C1-C2 and O1-H distances, colored by the Deep-LDA CV. The gray dashed lines correspond to the isolines of the free energy surface projected onto this space, while the black solid one corresponds to the minimum free energy path, computed using the nudged elastic band method \cite{JONSSON1998}. (d) Free energy surface projected along the C1-C2 distance, computed with a block average over the second part of the simulation, every 1 ns. The standard error is below 1 kJ/mol and is thus not visible.} 
  \label{fig:aldol}
\end{figure}

Having defined the input descriptors, we train the Deep-LDA network on unbiased simulations of the two states and use the resulting CV in combination with OPES to enhance the process (see SI-4 for the computational details). The time evolution of the Deep-LDA CV is reported in fig.~\ref{fig:aldol}a, which illustrates how the system is reversibly driven from reactants to products.

To assess the convergence of the simulation, we compute the FES along the C1-C2 distance (see inset of fig.~\ref{fig:aldol}c) with a block average and report the results in fig.~\ref{fig:aldol}d. The uncertainty is 0.25 kJ/mol in the regions of the metastable minima and about 1 kJ/mol close to the free energy barrier.

In order to gain a physical understanding of the process, we perform a feature importance analysis on the NN. More precisely, we rank the features by summing the modulus of the weights between the input and the first layer, multiplied by the standard deviation of the inputs in the training set. Another option is to calculate the derivatives of the Deep-LDA CV with respect to each input, averaged over all inputs in the training set. We found that these two ways of estimating feature relevance produce similar results (see fig. SI-4.3).
As shown in fig.~\ref{fig:aldol}b, there are three relevant descriptors, separated by a large gap from the following ones. These are the C1-C2 contact associated with the carbon-carbon bond together with the O1-H and O2-H contacts that characterize the proton transfer. This result has been obtained without any \textit{a priori} information on the reaction pathway and it is in agreement with chemical intuition. 

The FES projected on two of these variables has an easily understandable structure (fig.~\ref{fig:aldol}c and also fig. SI-4.2). The two basins are clearly separated and the Deep-LDA CV shows different values in the two basins. As in the alanine dipeptide example, the isolines of Deep-LDA resemble quite closely the ones of the FES. This implies that the direction along which the system is driven is correlated with the minimum free energy path, thus acting as a committor-like collective variable. 




\textbf{Conclusions}
We introduce in this letter a method that compresses the information from the metastable states into CVs. The method relies on a non-linear dimensionality reduction, performed by a NN, followed by a linear transformation. This is achieved by maximizing the LDA objective function, which searches for the representation that best separates the states. We show that the method effectively draws a path in the descriptors' space.
The Deep-LDA CV can be used in combination with any CV-based enhanced sampling method. We wish to remind the reader that our purpose here is not to find the ideal CV, but rather to build a variable that is good enough to promote transitions between the metastable states, given only a limited amount of information. We expect this approach to work best when the starting basins are adjacent in the phase space. If the system presents intermediate metastable states, they can be added iteratively as new classes for Deep-LDA, either by building a CV for each pair of states \cite{Rizzi2019} or by using a multi-class approach \cite{Piccini2018}. This CV could be further refined to follow the reaction pathway closely, either by including weighted data from biased simulations or by combining it with methods designed to extract the slowest relaxation modes \cite{McCarty2017c,Tiwary2016}. 
Another result is that the Deep-LDA features ranking can be used to identify the relevant descriptors in a data-driven way, as well as filtering them. For all these reasons we believe that our method could be of help in studying a large variety of rare events, including but not limited to chemical reactions, nucleation events and ligand-binding processes.

\section*{Acknowledgments}
The authors thank Sandro Bottaro, Michele Invernizzi and GiovanniMaria Piccini for carefully reading the paper, Manyi Yang for providing the inputs of the aldol reaction and Riccardo Capelli and Emanuele Grifoni for useful discussions.
This research was supported by the NCCR MARVEL, funded by the Swiss National Science Foundation, the Swiss National Science Foundation Grant No. 200021\_169429/1 and the European Union Grant No. ERC-2014-AdG-670227/VARMET.
The calculations were carried out on the Euler cluster of ETH Zurich.

\clearpage

\renewcommand{\thesection}{SI-\arabic{section}}
\renewcommand{\thesubsection}{SI-\arabic{subsection}}
\renewcommand{\thefigure}{SI-\arabic{figure}}
\renewcommand{\theequation}{SI-\arabic{equation}}
\renewcommand{\theequation}{SI-\arabic{equation}}
\emergencystretch=1em
\counterwithin{figure}{subsection}
\counterwithin{equation}{subsection}
\setcounter{figure}{0}  
\setcounter{equation}{0}  

\section*{SUPPORTING INFORMATION} 
\subsection{Neural network training}
\label{sec:S-nn}
The Deep-LDA model is trained using the PyTorch \cite{AdamPaszke;SamGross;etal2017} library. We report here the parameters used for the training of the neural network. We first apply a regularization to the within class scatter matrix in the form of $\mathbf{S}'_w=\mathbf{S}_w+\lambda\mathbf{I}$, with $\lambda=0.05$. 

The loss function is composed by three terms:
\begin{equation}
    \mathcal{L} = -v_1 \ -\ \alpha\ \frac{1}{1+\left( \overline{s^2}  -1 \right)^2}\ +\ \gamma \sum_i |\theta_i|^2  
\end{equation}
where the first one is the LDA eigenvalue of eq. 4, the second one is the regularization on the CVs values introduced in eq. 6, and the third one is an L2 regularization over the weights $\theta_i$ of the network, with  $\gamma=10^{-5}$. We found that the intensity $\alpha$ of the regularization term is connected to the value of $\lambda$ in $\mathbf{S}'_w$, and a robust behavior is obtained when their product is kept constant. In fact, $\alpha$ affects the numerator of Fisher's ratio (eq. 3), while the latter only the denominator. Thereby, we use $\alpha=2/\lambda$.
We use the rectified linear unit as activation function. The training data, which consists of 10000 configurations, is divided into batches of size 2000, and the model is optimized using ADAM \cite{Kingma2014} with a learning rate of $10^{-4}$, until the loss function converges. In order to prevent overfitting, a small set of configurations from different simulations is used as a validation set, and the early stopping technique is used where necessary. Once the model has been trained, the LDA coefficients are obtained using the whole training set and the model is exported.

\subsection{PLUMED interface with PyTorch}
\label{sec:S-plumed}
A development version of the open-source plug-in PLUMED2 \cite{Tribello2014} is used. At first, we define the descriptors which form the input for the Deep-LDA network. The NN architecture and parameters are loaded using a custom interface via LibTorch C++ APIs. Finally, the Deep-LDA collective variable is used in combination with the OPES method to enhance the sampling.

\subsection{Alanine dipeptide}
\label{sec:S-ala2}
\textbf{Computational details.} 
The alanine dipeptide simulations are carried out using GROMACS 2019.4 \cite{VanDerSpoel2005a} patched with PLUMED 2.5. We use the Amber99-SB force field \cite{Hornak2006} with a time step of 2 fs. The NVT ensemble is sampled using the velocity rescaling thermostat \cite{Bussi2007} with a temperature of 300K. For OPES we use \verb|SIGMA=0.05| and \verb|BARRIER=30 kJ/mol|. The neural network used to model the Deep-LDA CV has 3 hidden layers with \{30, 15, 5\} neurons per layers. The distances are standardized to take values from -1 to 1.

\begin{figure}[h]
  \centering
  \includegraphics[width=0.9\columnwidth]{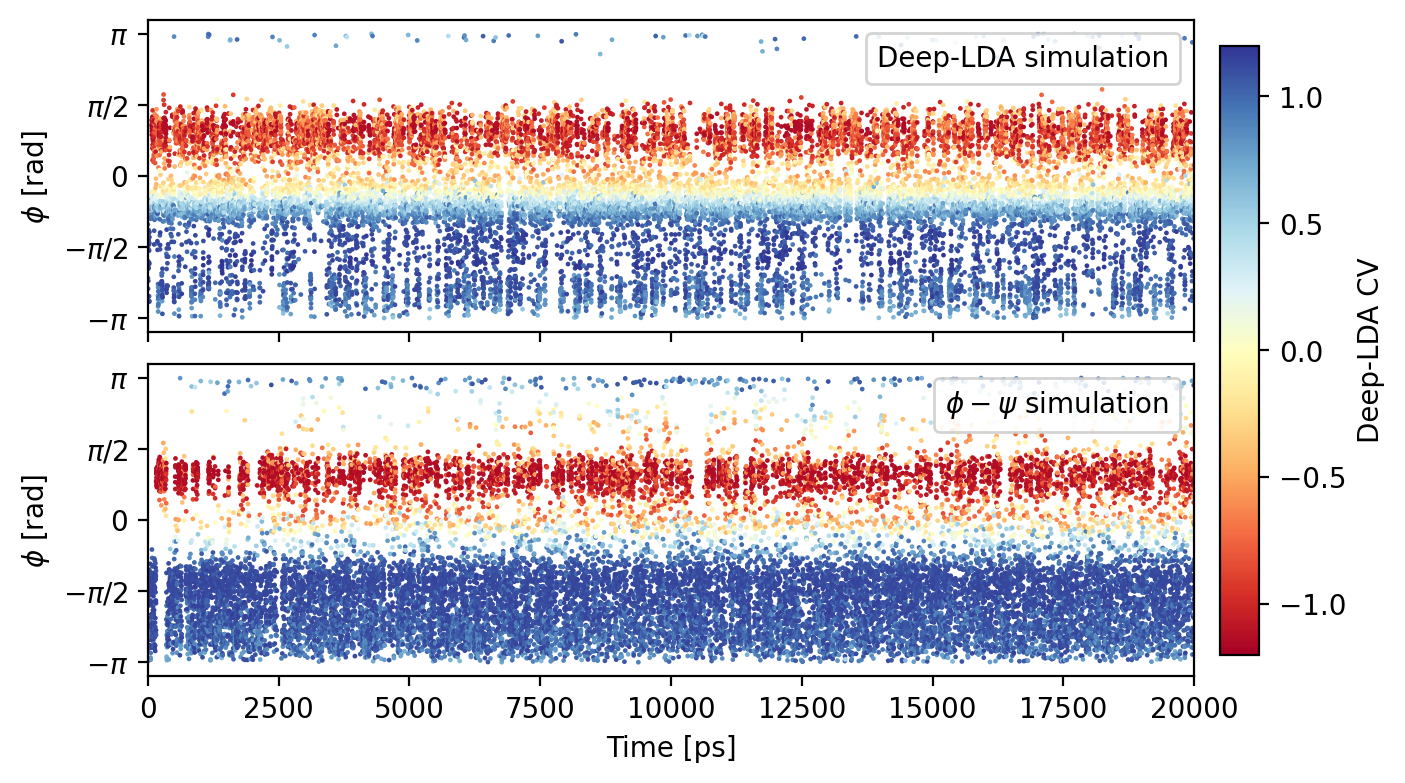}
  \caption{Time evolution of the $\phi$ angle in the Deep-LDA OPES simulation (top) and the optimal case when the two dihedral angles are used to define the bias potential (bottom).}
  \label{fig:S-ala-compare}
\end{figure}

\bigskip
\textbf{Comparison with $\phi$ and $\psi$.}
In fig. \ref{fig:S-ala-compare} we report the trajectory of a simulation using Deep-LDA, compared with the case in which the two Ramachandran angles are used. As it can be seen from the transition rate, despite the fact that only scalar distances are used as inputs for Deep-LDA, the NN manages to build a non-linear combination of them which is able to distinguish and to drive transitions between the states, thus mimicking the effect of the dihedral angles.  

\begin{figure}[h]
 \centering
        \includegraphics[width=0.6\columnwidth]{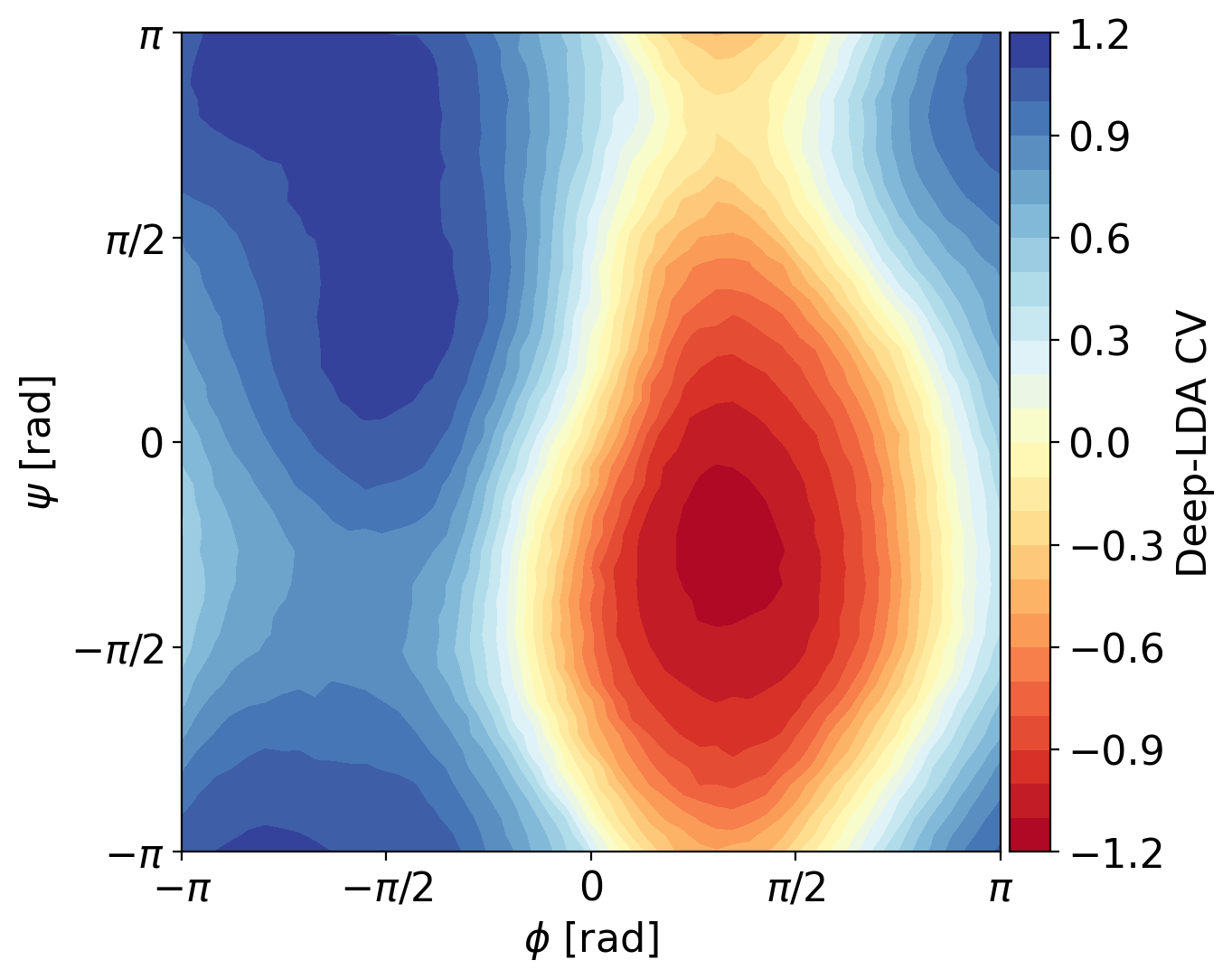} 
        \caption{Probability distribution of the Deep-LDA CV in the Ramachandran plot.}
      \label{fig:S-ala-isolines}
\end{figure}

\bigskip
\textbf{Conditional probability distribution of the Deep-LDA CV in the Ramachandran plot.}
In fig. \ref{fig:S-ala-isolines} we report the probability distribution of the Deep-LDA CV as a function of the two angles, $p(s\,|\,\phi,\psi)$. This has been computed from a simulation with OPES and an uniform target distribution in the \{$\phi$,$\psi$\} space.
The Deep-LDA CV is able to reproduce the topology of the free energy landscape of alanine dipeptide, while using only short unbiased simulations.

\begin{figure}[h]
  \centering
  \includegraphics[width=.95\columnwidth]{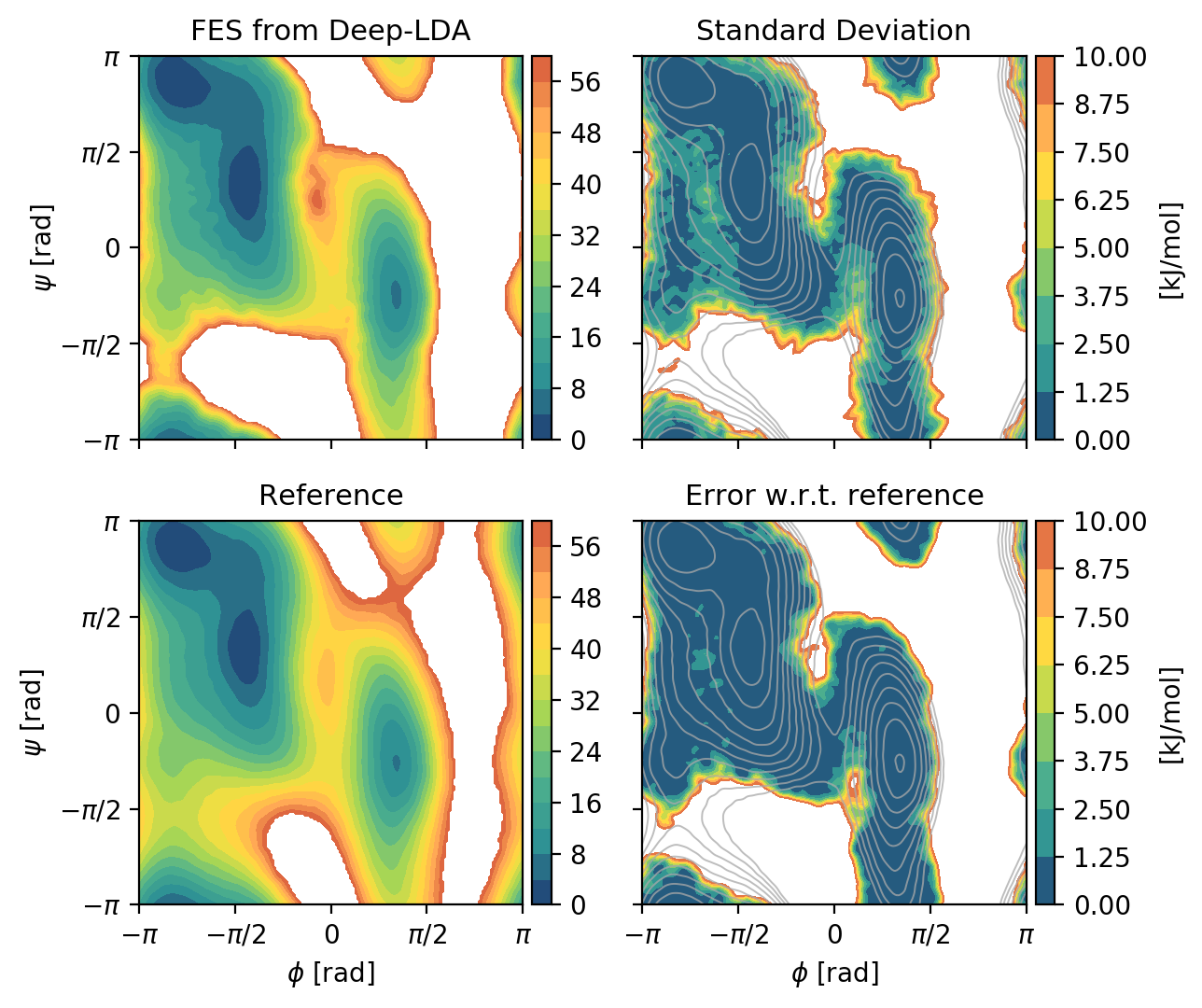}
  \caption{Free energy surface of alanine dipeptide in the $\phi$,$\psi$ space. We show the free energy mean value reweighted from 10 OPES simulations where the Deep-LDA CV was biased (top right) and its corresponding standard deviation (top left). We then show a reference free energy obtained by biasing directly the two dihedrals $\phi$ and $\psi$ (bottom left) and the absolute error between the reference and the Deep-LDA simulations (bottom right). In the panels on the right hand side, we display the isolines of the FES to highlight the position of the the metastable states.}
  \label{fig:S-ala-fes-2d}
\end{figure}

\bigskip
\textbf{Accuracy of the reweighted landscape.}
We quantify the accuracy of the reconstructed free energy landscape in terms of the physical variables, $\phi$ and $\psi$. The FES can be obtained from OPES simulations via an umbrella-sampling reweighting \cite{Invernizzi2019}:
\begin{equation}
    P(\mathbf{s})=\frac{\langle \delta\left[\mathbf{s}-\mathbf{s}(\mathbf{R})\right]\ e^{\beta V(\mathbf{s})} \rangle_V}{\langle e^{\beta V(\mathbf{s})} \rangle_V}
\end{equation}
In figure \ref{fig:S-ala-fes-2d} we report the results obtained from 10 independent simulations, which we use to quantify the standard deviation, as an estimate of the error of a single simulation. Furthermore, for a direct comparison, we report the reference FES obtained when biasing directly the two dihedral angles, from which we estimate the absolute error with respect to the reference. The error made in reconstructing the FES is smaller than 0.5 $\mathrm{k_B T}$ (1.25 kJ/mol) in the most relevant regions. Moreover, the standard deviation is smaller than 1 $\mathrm{k_B T}$  (2.5 kJ/mol) in all the regions of interest.

\begin{figure}[h]
  \centering
  \includegraphics[width=.8\columnwidth]{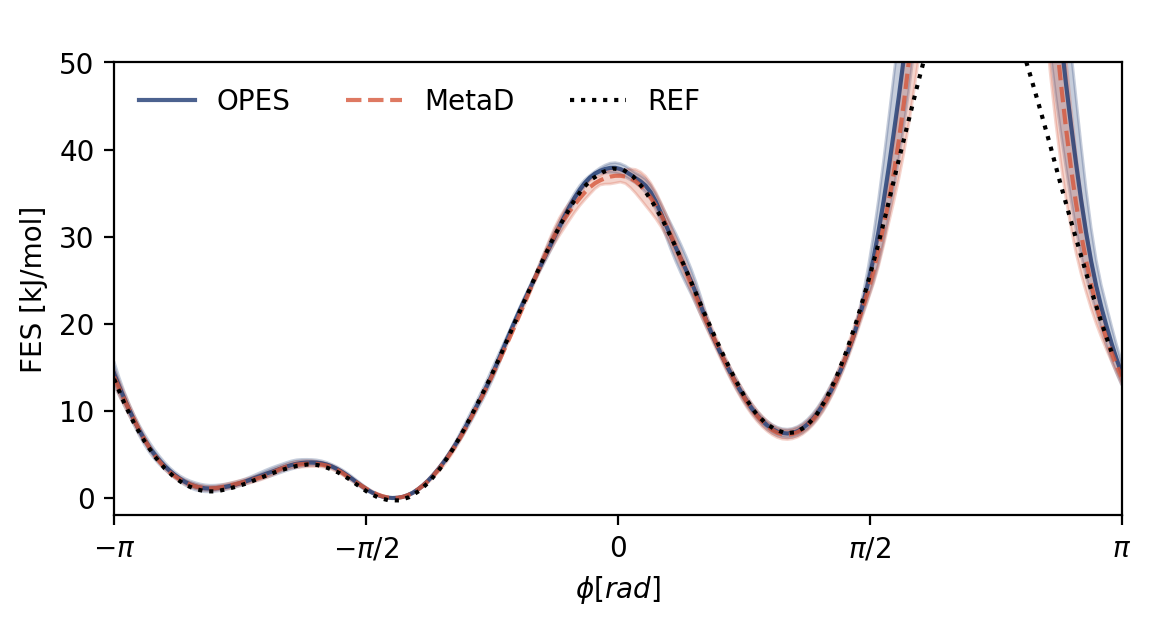}
  \caption{Free energy surface along the $\phi$ variable from simulations with OPES and Well-Tempered Metadynamics where the Deep-LDA CV was biased. For comparison, the reference free energy value is shown.}
  \label{fig:S-ala-fes-meta}
\end{figure}

\clearpage
\bigskip
\textbf{Metadynamics simulations.}
Here we combine the Deep-LDA CV with Well-Tempered Metadynamics \cite{Barducci2008}.
We follow the same protocol as above, running 10 independent simulations to calculate the mean and the standard deviation of the free energy. The Metadynamics parameters are \verb|SIGMA=0.05|, \verb|GAMMA=6| and \verb|PACE=500|. In fig.~\ref{fig:S-ala-fes-meta} we show the FES along the $\phi$ variable. 

\begin{figure}[h]
 \centering
        \includegraphics[width=0.9\textwidth]{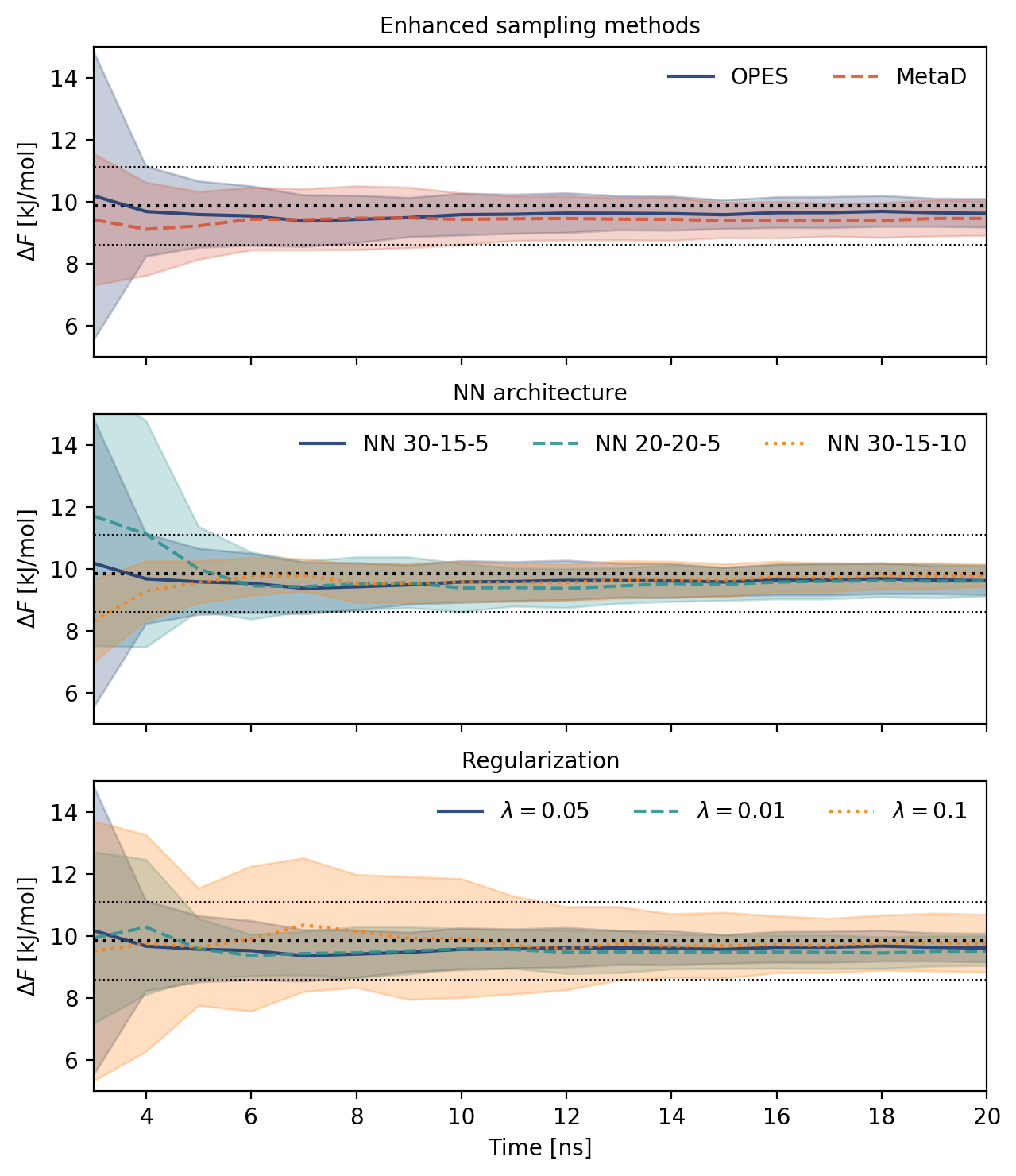} 
        \caption{Time evolution of the free energy difference for alanine dipeptide in a number of different simulations. 
        The mean (solid line) and standard deviation (shadow area) over 10 simulations are reported, together with the reference value (thicker dotted line) and a region of 0.5$\mathrm{k_B T}$ from within the reference (thinner dotted lines).
        In the top panel we compare OPES and Metadynamics simulations.
        In the central panel, we show the effect of different NN architecture.
        In the bottom panel, we vary the regulatization parameter $\lambda$ (and therefore also $\alpha$ as we always keep $\alpha = 2/\lambda$).}
      \label{fig:S-ala-convergence}
\end{figure}

\bigskip
\textbf{Convergence tests.}
We provide a test of the robustness of the Deep-LDA CV with respect to different enhanced sampling methods and different realizations of the NN by assessing the convergence of the alanine dipeptide free energy difference.
After training different Deep-LDA CVs (by changing only one parameter at the time), our protocol is to run 10 different simulations for each case and calculate mean value and the standard deviation of the free energy difference. The free energy difference between the basins is defined as follows:
\begin{equation}
\Delta F = \frac{1}{\beta} \log \frac{\int_A e^{-\beta F(\phi)} \mathrm{d} \phi }{\int_B e^{-\beta F(\phi)} \mathrm{d}\phi} 
\end{equation}
where $\phi$ is the dihedral angle and $F(\phi)$ is the reweighted free energy. The two integrals are computed over the regions of phase space corresponding to metastable state A and B, in this case $\phi<0$ and $\phi>0$. The first 2 ns of the simulation are discarded and the update of $\Delta F$ is performed every 1 ns. 
In fig.~\ref{fig:S-ala-convergence}, we compare the time evolution of the free energy difference in the different cases.
All examples show a quick convergence of $\Delta F$ within the 0.5 $\mathrm{k_B T}$ uncertainty.

\clearpage

\subsection{Aldol reaction}
\label{sec:S-aldol}
\begin{figure}[h]
  \centering
  \includegraphics[width=0.25\columnwidth]{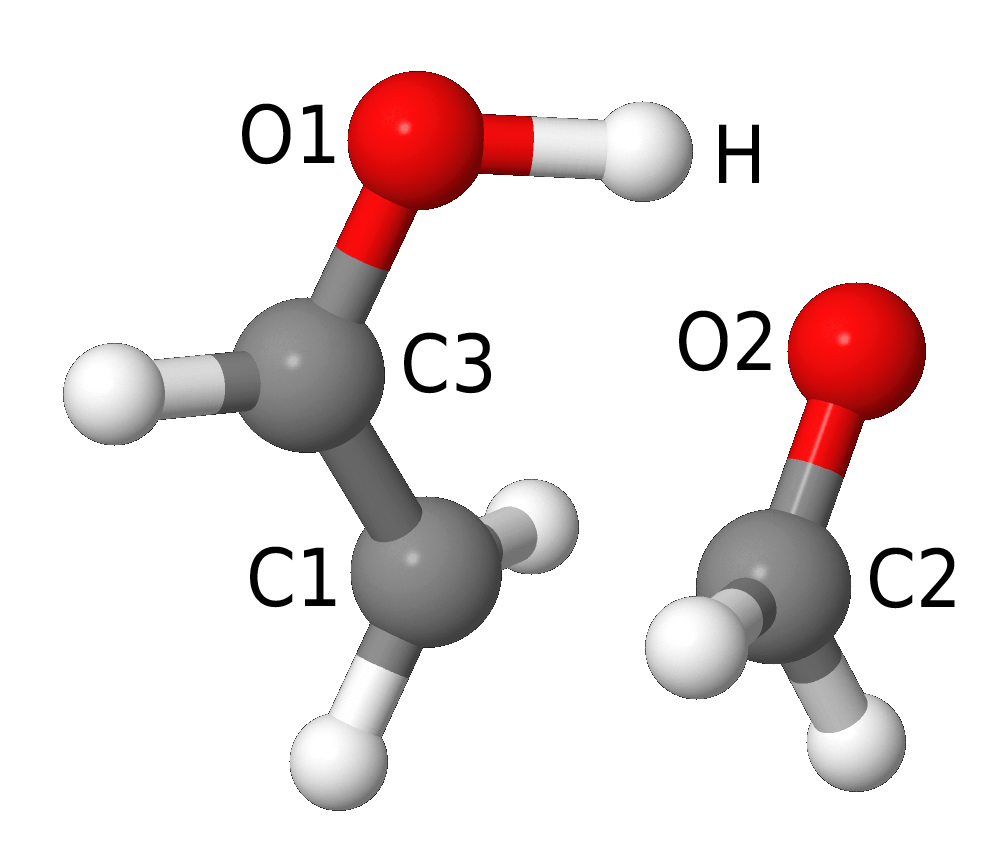}
  \caption{Snapshot of the aldol reaction with labels on the relevant atoms involved in the process. The C1-C2 bond has to be formed, with a simultaneous proton transfer from O1 to O2.}
  \label{fig:S-aldol-labels}
\end{figure}

\textbf{Computational details.}
The software package CP2K 7.1 \cite{Hutter2014} is used to carry out the simulations of the aldol chemical reaction at the PM6 semi-empirical level. We used an integration step of 0.5 fs, employing the same thermostat as in \ref{sec:S-ala2} with a time constant of 100 fs. In fig.~\ref{fig:S-aldol-labels} we report a snapshot of the aldol reaction with the labeling of the atoms used in the paper. The contacts between the atoms are computed using PLUMED, with the parameter $\sigma_{ij}$ of the switching function of eq. 10 being equal to 1.7 \r{A}  for C-C, 1.6 \r{A} for O-O and C-O, and 1.2 \r{A} for C-H and O-H species. The exponents of the contact functions are chosen as $n=6$ and $m=8$ to enforce a smooth behavior over a wide range of distances. The whole set is made of 40 contacts. A neural network with three hidden layers and \{24, 10, 4\} nodes per layer is used. In OPES, we use \verb|SIGMA=0.1| and \verb|BARRIER=160 kJ/mol|. 
The Deep-LDA CV $s$ is very sharp in the metastable basins, thus it is not ideal for usage in enhanced sampling. To broaden it, we transform it by $s'=s+s^3$ and use OPES to enhance the dynamics of $s'$. A restraint $(k/2)(s'-s_0)^2$ with $k=2000$ kJ/mol and $s_0=\pm 3.2$ is put to prevent $s'$ to explore regions beyond its scope. To set a limit on the mutual distance between the vinyl alcohol and the formaldehyde, we also put a restraint on the distance $d$ between their respective centers of mass of the form $k(d-d_0)^2$ with $k=150$ kJ/mol and $d_0 = 5$ \r{A}.

\begin{figure}[h]
  \centering
  \includegraphics[width=\columnwidth]{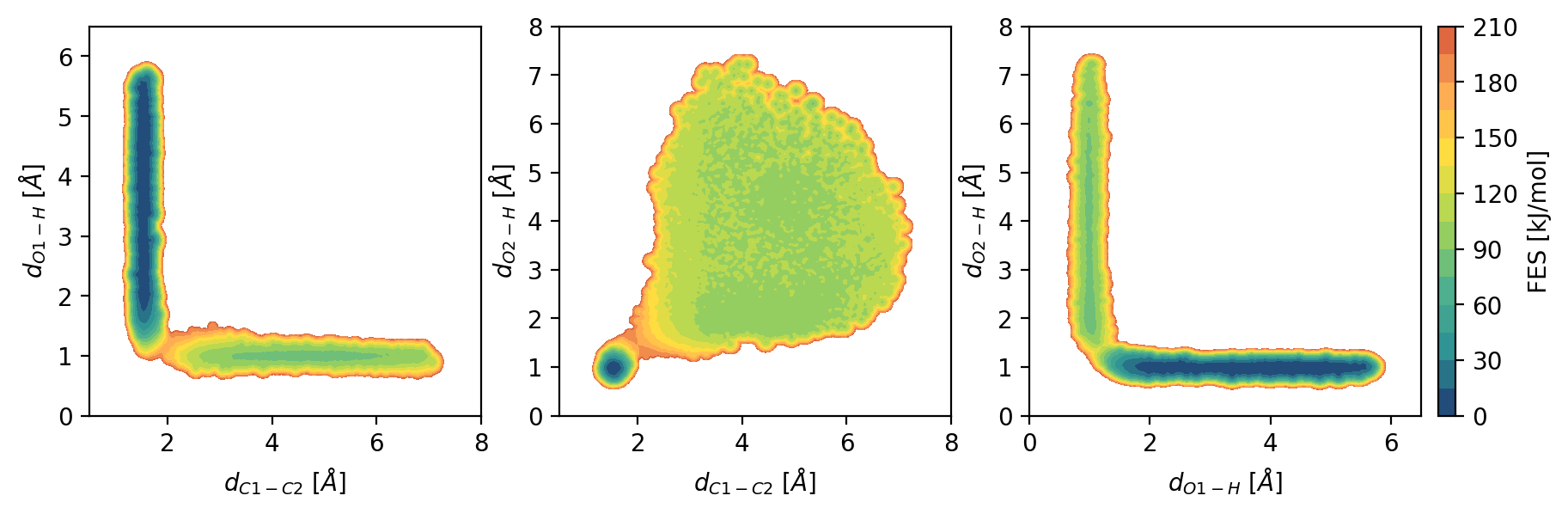}
  \caption{Free energy surface projected along combinations of the three chemical distances identified as the most important from the Deep-LDA ranking.}
  \label{fig:S-aldol-fes}
\end{figure}

\textbf{Free energy landscape projected onto chemical distances.}
We report in fig. \ref{fig:S-aldol-fes} the free energy profiles obtained via reweighting of the Deep-LDA CV in the space of the three chemical distances identified by the features importance analysis.

\begin{figure}[h]
  \centering
  \includegraphics[width=0.5\columnwidth]{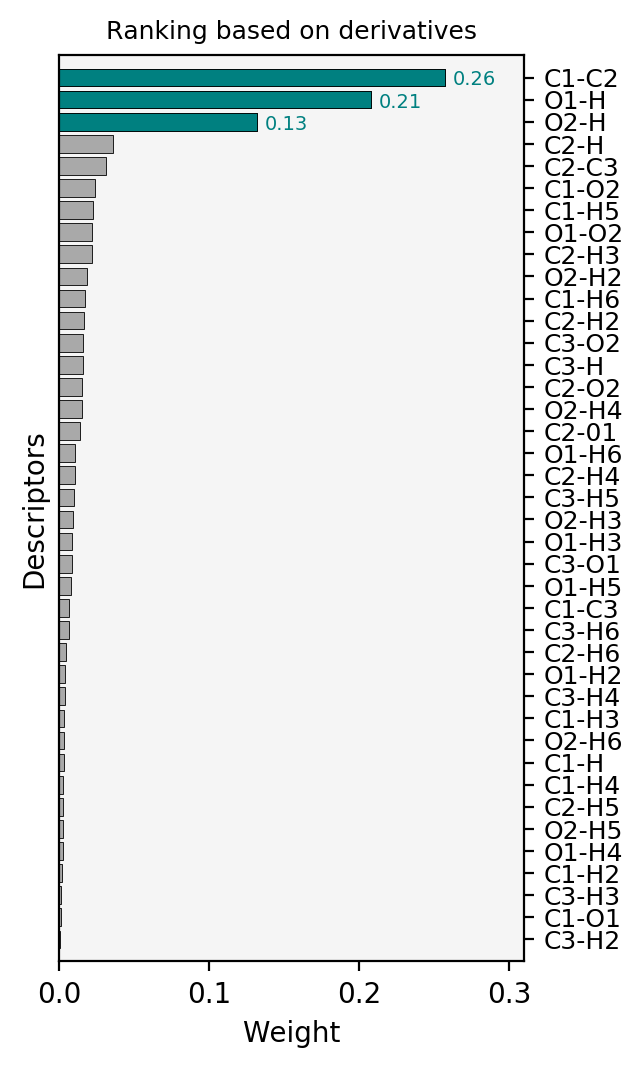}
  \caption{Features ranking based on the derivatives of the Deep-LDA CV with respect to inputs.}
  \label{fig:S-aldol-ranking}
\end{figure}

\clearpage
\bigskip
\textbf{Features ranking based on the derivatives}
We propose an alternative ranking of the input contacts which is based on the derivatives of the CV with respect to the inputs. The ranking for the k-th descriptor is defined as $r^k = \sum_{j=1}^{n} \, \left| \frac{\partial \, s_j}{\partial \, x^k_j} \right| \sigma(x^k)$ where the sum is performed over a set of $n$ configurations from the training set and $\sigma(x^k)$ is the standard deviation of descriptor $x^k$ over that set. The rankings are normalized so that their sum is equal to one. The resulting features ranking in fig~\ref{fig:S-aldol-ranking} is in qualitative agreement with the one proposed in fig. 4 as the three most relevant contacts correspond. 

\begin{figure}[h]
  \centering
  \includegraphics[width=0.8\columnwidth]{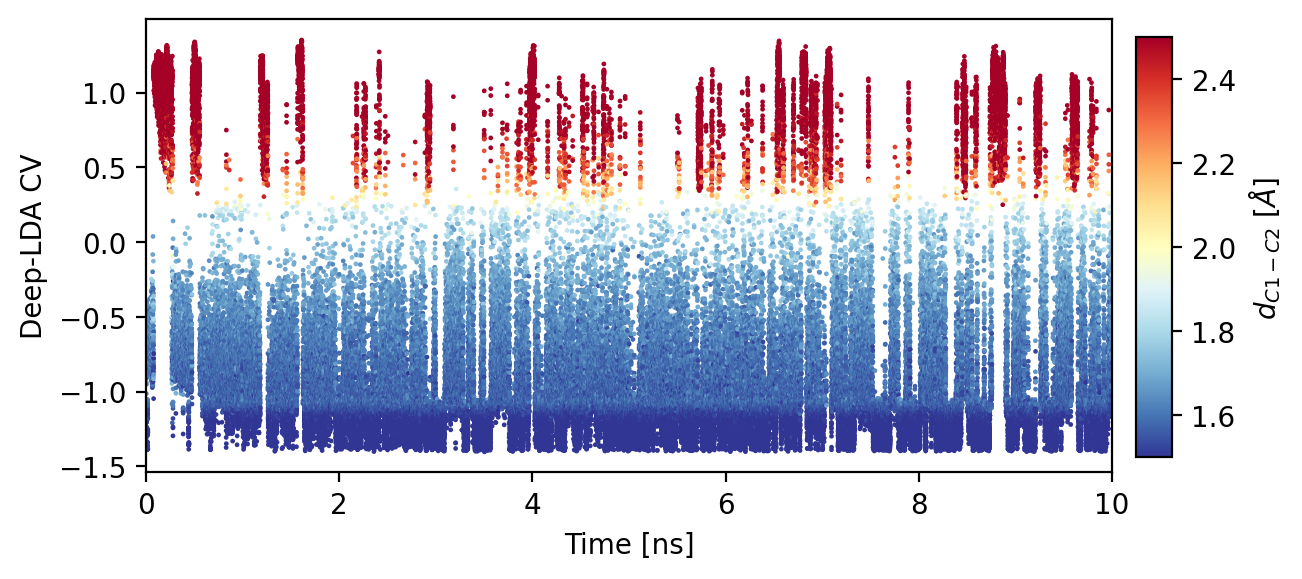}
  \caption{Time evolution of the Deep-LDA variable trained over all the distances of the aldol reaction. The color corresponds to the C1-C2 distance.}
  \label{fig:S-aldol-compare}
\end{figure}

\clearpage
\textbf{Distances as inputs for the Deep-LDA CV}
We train a Deep-LDA CV using all the distances instead of the contacts. In analogy with the simulations above, we transform it with $s'=s+s^3$ and feed it to OPES with \verb|SIGMA=0.09| and \verb|BARRIER=160 kJ/mol|. The resulting CV dynamics is shown in fig.~\ref{fig:S-aldol-compare}. The CV is able to drive the system back and forth between the basins, albeit in a slightly less efficient way than the CV trained with the contacts. 

\subsection{Data availability}
All the code and the input files needed to reproduce the results reported in this paper are openly available in the PLUMED-NEST repository (www.plumed-nest.org), as plumID:20.004. The results are deposited in the Materials Cloud Archive (www.materialscloud.org), with id: 2020.0035.

\clearpage


\printbibliography

\end{document}